\documentclass[12pt]{article} 

\usepackage[doublespacing]{setspace}

\usepackage[numbers,round,sort&compress]{natbib}

\bibpunct{\textit{\textbf{(}}} {\textit{\textbf{)}}}{,}{n}{}{;}

\usepackage{amssymb} 
\usepackage{dsfont}
\usepackage{amsmath}
\usepackage{amsfonts}
\usepackage{latexsym}
\usepackage{placeins}
\usepackage{mathptmx}       
\usepackage{helvet}         
\usepackage{courier}        
\usepackage{type1cm}        
\usepackage{makeidx}         
\usepackage{graphicx}        
\usepackage{multicol}        
\usepackage[bottom]{footmisc}
\usepackage{pdflscape}
\usepackage{epstopdf} 

\usepackage{courier}
\usepackage{listings}
\usepackage{color}
\usepackage{xcolor}
\usepackage{footnote}
\usepackage[bottom]{footmisc}   
\usepackage{url}
\usepackage{upquote}
\usepackage{algorithm2e}
\usepackage{wrapfig}
\usepackage{tikz}

\definecolor{dkgreen}{rgb}{0,0.6,0}
\definecolor{edviolet}{rgb}{0.57,0.32,0.74}
\definecolor{gray}{rgb}{0.5,0.5,0.5}
\definecolor{light-gray}{gray}{0.97}

\usepackage{hyperref}
\hypersetup{colorlinks=true,allcolors=edviolet}
\usepackage{breakurl}

\lstdefinelanguage{rule}
    {morekeywords={\^, \%agent:, \~, \%token:, \%var:, ->, <->, rbmo:Agent, rbmo:Rule, rbmo:Site,rbmo:State, rbmo:Site, rbmo:hasSite, rbmo:hasState, rbmo:stateAssignment, rbmo:hasSubrule, rbmo:Pattern, rbmo:Expression, rbmo:Kappa, rbmo:agent, rbmo:state, rbmo:site, rbmo:BioNetGen, SBO:0000002, rbmo:internalState, rbmo:isStatusOf, rbmo:BoundState, rbmo:UnboundState, rbmo:status, rbmo:lhs, rbmo:rhs, rbmo:isBoundBy, rbmo:Observable},
     basicstyle=\fontsize{7}{9}\selectfont\ttfamily,
     backgroundcolor=\color{white},
     keywordstyle=\color{blue},
     commentstyle=\color{gray},
     stringstyle=\color{dkgreen},
     tabsize=2,
     showspaces=false,
     showstringspaces=false,
     breaklines=true,                           
     sensitive=true,                            
     morestring=[b]",                           
     escapeinside={$}{$},
     alsoletter=\%:-\textgreater\textless\^\<\>
}

\newcommand{\lstsetrule}{
 \lstset{language=rule,
        frame=single,
        tabsize=2,
        xrightmargin=4pt,xleftmargin=4pt,
        framesep=4pt,
        aboveskip=0pt,
        belowskip=0pt
 }
}

\newcommand{\rbmterm}[1]{\texttt{\footnotesize\color{blue}#1}}
\newcommand{\supplementary}[1]{}

\begin{document}


\LARGE
\noindent \textbf{Annotations for Rule-Based Models} 

\large
\noindent \textbf{Matteo Cavaliere,  Vincent Danos, Ricardo Honorato-Zimmer and William Waites}

\normalsize

\noindent All authors contributed equally. \\

\noindent \textbf{Corresponding author}: \\
William Waites \\
Laboratory for Foundations of Computer Science \\
School of Informatics \\
University of Edinburgh \\
Edinburgh, EH8 9LE, UK \\
Email: \href{mailto:wwaites@inf.ed.ac.uk}{\texttt{wwaites@inf.ed.ac.uk}}

\par\vspace{2\baselineskip}\noindent \small{This manuscript has been prepared for inclusion in
\textit{Modeling Biomolecular Site Dynamics: Methods and Protocols} (Ed. William S. Hlavacek),
part of the \href{https://www.springer.com/series/7651}{\textit{Methods in Molecular Biology}}
series.

\clearpage

\section*{Summary}
 The chapter reviews the syntax to store machine-readable annotations and describes the mapping between rule-based modelling entities (e.g., agents and rules) and these annotations.
 In particular, we review an annotation framework and the associated guidelines for annotating rule-based models, encoded in the commonly used Kappa and BioNetGen languages, and present prototypes that can be used to extract and query the annotations.
 An ontology is used to annotate models and facilitate their description.
 \\

\noindent \textbf{Key words}: Rule-Based Modelling, Kappa, BNGL, KaSim, BioNetGen, RDF, Turtle, MIRIAM, SPARQL, Rule-Based Model Ontology (rbmo)

\clearpage

\section{Introduction}

\subsection{The need for model annotation}

The last decade has seen a rapid growth in the number of model
repositories~\cite{Li2010,Yu2011,Snoep2003,Misirli2014,Moraru2008}. It is also well understood that the creation of models and of repositories requires expert
knowledge and integration of different types of biological data from
multiple sources~\cite{Endler2009}. These data are used to derive the
structure of, and parameters for, models.
However which data are used and how the model is derived
from that data is not part of the model unless we explicitly annotate it in a well-defined way.

In general, annotations decorate a model with metadata linking to biologically relevant information~\cite{Blinov2010}. Annotations can facilitate the automated exchange, reuse and composition of complex models from simpler ones.  Annotations can also be used to aid in the computational conversion of models into a variety of other data
formats. For example, PDF documents~\cite{Li2010} or visual
graphs~\cite{Funahashi2007} can be automatically generated from
annotated models to aid human understanding.

On the computational and modelling side, rule-based languages such as
Kappa \cite{Danos2004,DanosFFK07} and the BioNetGen language (BNGL) \cite{Faeder2009}
have emerged as helpful tools for modelling biological systems~\cite{KohlerKV14}.
One of the key benefits of these languages is that they can be used to
concisely represent the combinatorially complex state space
inherent in biological systems. Rule-based modelling languages have facilities to add comments that are intended for unstructured documentation and usually directed at the modeller or
programmer. These comments are in general human and not machine-readable. 
This can be a problem because the biological semantics of the model entities are not computationally accessible and cannot be used to influence the processing of models.

Previous works have addressed the issue of annotations in rule-based models. In particular, Chylek et al. \cite{Chylek2011} suggested extending rule-based
models to include metadata, focusing on documenting models with
biological information using comments to aid the understanding of
models for humans. More recently, Klement et al. \cite{Klement2014} have
presented a way to add data in the form of property/value pairs using
a specific syntax. On the other hand, machine-readable annotations
have been applied to rule-based models using PySB, a programming
framework for writing rules using Python~\cite{Lopez2013}.  However,
this approach is restricted as annotations cannot be applied to sites or states.

In this chapter we first discuss the general idea of annotation, its relation with the concept of abstraction and then review an annotation framework for rule-based models that has recently introduced and defined by Misirli et al. \cite{Misirli2015annottation}.

\subsection{Reactions, rules, annotations and abstractions}

Before entering into the technicalities of the annotation
framework of interest, we would like to discuss in an informal and intuitive manner the differences between models created using reactions versus
those obtained using rules, discussing the advantages of considering
annotations and how they are strictly linked to the much more general
notion of abstraction. 

\subsubsection{Reactions and rules}

Rules as they are to be understood in the present context are a sort of generalisation of reactions of the type familiar from chemistry. The
reason this generalisation is useful can be easily seen. Consider the
following toy example,
\begin{equation}
  \tikz { \draw (0,0) circle (0.1); } + \tikz { \draw (0,0) circle (0.1); }
  \rightarrow
  \tikz {
    \draw (-0.2,0) circle (0.1);
    \draw (0.2,0) circle (0.1);
    \draw (-0.1,0) -- (0.1,0);
  }\nonumber
\end{equation}
which can be understood as a step in the creation of a
polymer from two monomers. Multiple applications of this rule result
in a progressively longer chain of molecules,
\begin{align*}
  \tikz {
    \draw (-0.2,0) circle (0.1);
    \draw (0.2,0) circle (0.1);
    \draw (-0.1,0) -- (0.1,0);
  } +
  \tikz { \draw (0,0) circle (0.1); }
  &\rightarrow
  \tikz {
    \draw (-0.4,0) circle (0.1);
    \draw (0,0) circle (0.1);
    \draw (0.4,0) circle (0.1);
    \draw (-0.3,0) -- (-0.1,0);
    \draw (0.1,0) -- (0.3,0);
  }\\
  \tikz {
    \draw (-0.4,0) circle (0.1);
    \draw (0,0) circle (0.1);
    \draw (0.4,0) circle (0.1);
    \draw (-0.3,0) -- (-0.1,0);
    \draw (0.1,0) -- (0.3,0);
  } +
  \tikz { \draw (0,0) circle (0.1); }
  &\rightarrow
  \tikz {
    \draw (-0.6,0) circle (0.1);
    \draw (0.6,0) circle (0.1);
    \draw (-0.2,0) circle (0.1);
    \draw (0.2,0) circle (0.1);
    \draw (-0.1,0) -- (0.1,0);
    \draw (-0.3,0) -- (-0.5,0);
    \draw (0.3,0) -- (0.5,0);
  }\\
  &\cdots
\end{align*}
Writing this down in the notation of reaction, we would need
to explicitly generate the entire unbounded sequence of
reactions with an unbounded number of chemical species,
\begin{align*}
  A + A &\rightarrow A_2\\
  A_2 + A &\rightarrow A_3\\
  A_2 + A_2 &\rightarrow A_4\\
  A_3 + A &\rightarrow A_4\\
  &\cdots
\end{align*}
Clearly this is unworkable with finite resources. The solution is to allow a
species to have \emph{sites} at which connections can be made. In the
above example, the species could be described as \texttt{A(u,d)},
that is substance \texttt{A} with an upstream and a downstream
site. The interaction can then be written as,
\begin{align*}
  \mathtt{A(d)},\; \mathtt{A(u)} \rightarrow \mathtt{A(d!1)},\; \mathtt{A(u!1)}
\end{align*}
where the notation \texttt{d} means that the downstream site is
unbound, and the \texttt{d!1} means it is bound with a particular edge. Note
that this says nothing about the state of the upstream site in the
first instance of \texttt{A} nor the downstream site in the second, so
there can be an arbitrarily long chain of molecules attached at
those sites. It is easy to see that this compact notation captures
both the infinite sequence of reactions and the infinite set of
species that would be required to express the same interaction as
a set of chemical reactions.

\subsubsection{Annotations}

Informally, the word ``annotation'' has a meaning similar to
``documentation'' but with a difference in specificity. Whereas
documentation connotes a rather large text describing something (e.g., an object),
annotation is expected to be much shorter. It also evokes proximity:
it should be in some sense ``near'' or ``on'' the thing being
annotated. In both cases there seems to be a sharp distinction between
the text and its object. The object should exist in its own right, be
operational or functional in the appropriate sense without need to
refer to exogenous information. Annotation might help to understand the
object but the object exists and functions on its own.

This folk theory of annotation breaks down almost immediately under
inspection. A typical example is data about a book such as might
be found in a library catalogue. This is a canonical
example used to explain what is meant by \emph{metadata} or data about
data. The first observation is that if we look at a book and peruse
the first few pages it is almost certain that we will find information
about who wrote it and where and when it was published. This
information is not the book, it is metadata about the book, but it is
contained within the covers of the book itself.

Perhaps this is not so serious a problem. It is possible in principle
to imagine that a book, say with the cover and first few pages torn
out, is still a book that can be read and enjoyed. Perhaps somehow the
metadata is \emph{separable} and that is the important idea. The
book-object can exist on its own and serve its
purpose independently of any annotation or metadata. While the metadata
might usually be found attached to the book, can easily be removed without
affecting the fundamental nature of the book itself.

But what of other
things that we might want to do with a book?
A favourite activity of academics is \emph{citing} documents such as
books and journal articles. This means including enough information in
one work to unambiguously \emph{refer} to another. There is an urban
legend that Robarts Library at the University of Toronto is said to be
sinking because the engineers charged with building it did not account
for the weight of the books within. Supposing that this were true,
these poor apocryphal engineers could have used metadata within the
university's catalogue to sum up the number of pages of all the books
and estimate their weight to prevent this tragedy. This summing is a computation
that operates purely on the metadata and not on the books themselves.

More mundanely, categorising and counting books in order to plan for
the use of shelf space in a growing collection, or even locating a
book in a vast library seem to be a plausible things to do with
metadata that do not involve any actual books.  Manipulation and
productive use of annotation is possible in the absence of the objects
and well-defined even if the objects no longer exist. One imagines
the despondent librarians and archivists of Alexandria making such
lists to document and take stock of their losses after the great
fire.

Now suppose that this list created by the librarians of Alexandria
itself ended up in a collection in some other library or museum. It is
given a catalogue number, the year it was acquired is marked. Now what
was metadata has now itself become the object of annotation! Here
we arrive at the important insight: what is to be considered
annotation and what is to be considered object depends on the purpose
one has in mind. If the interest is the collection of books in
Alexandria, the list is metadata, a collection of annotations, about
them. If the interest is in the documents held by a contemporary
museum, among which the list is to be found, the list is an
object. The distinction is not intrinsic to the objects themselves.

Turning to the subject at hand, the objects to be annotated are
rules. According to the folk theory of annotation, there should be a
sharp distinction between rules and their annotation. When it
comes to executing a simulation, the software that does this need not
be aware of the annotations. Indeed the syntax for annotating rules
described here is specifically designed for backwards compatibility
such that the presence of annotations should not require any
disruption or changes to existing simulation software.

So long as the \emph{purpose} of the annotations is as an aid to
understanding the rules the location of the distinction between rule
and annotation is fixed in this way. The obvious question is, are
there other uses to which the annotations can be put? 

In the report of Misirli et al. \cite{Misirli2015annottation}, where the annotation mechanism of interest was first described, one of the
motivating examples was to create a \emph{contact map}, a type of
diagram that shows which agents or species interact with each other
and labels these interactions with the rule(s) implementing them (an example of a contact map is provided later in this chapter).

Use of a contact map is illustrative of how movable the separation
between object and annotation is~\cite{buneman2013annotations}. The
entities of interest, rules and agents, are on the one hand decorated
with what seems to be purely metadata: labels, or friendly
human-readable names that are suitable for placing on a diagram,
preferable to the arbitrary machine-readable tokens that are used by
the simulator (arbitrary because they are subject to renaming
as required). On the other hand, the interactions between the
substances, what we wish to make a diagram \emph{of}, are written down
in a completely different language with an incompatible syntax.

A minor change of perspective neatly solves this problem. It is simply
to rephrase the rule, saying ``A and B are related, and the way they
are related is that they combine to form C''. This has the character
of annotation: the rule itself is a statement about the substances
involved. More particularly it describes a \emph{relation} between the
substances. On close inspection, giving a token used in a rule a
human-readable name is also articulating a relation, that is the
relation called ``naming'' between the substance and a string of
characters suitable for human consumption.

With this change of perspective, all of the information required to
make the diagram is now of the same kind. The only construct that
must be manipulated is sets of relations between entities (and strings
of text, which are themselves a kind of entity). Fortunately there
exist tools and query languages for operating on data stored in just
this form. Having worked out the correct query to extract precisely
what is needed to produce the diagram, actually generating it is trivial.

\subsubsection{Abstractions and annotations}

The preceding section on annotation, describes what can be thought of as a ``movable
line''. ``Above'' this line are annotations and ``below'' it are the
objects. The sketch of a procedure for producing a diagram to help
humans understand something about a system of rules as a whole
illustrated that it can be convenient to place this line somewhere
other than might be obvious at first glance --- and this example will
be considered in more detail below to demonstrate how this happens in
practice. However the idea of such a line and how it might be moved
and what exactly that means is still rather vague. Let us now make
this notion more precise.

Formally, a \emph{relation} between two sets, $X$ and $Y$, is a subset
of their Cartesian product, $X \times Y$. In other words it is a set
of pairs, $\left\{ (x, y)\, |\, x \in X, y \in Y\right\}$, and it is
usually the case that it is a proper subset in that not all possible
pairs are present in the relation. In order to compute with
relations, the sets must be \emph{symbols},
$X,\, Y \subseteq \mathbb{S}$, ultimately realised as sequences of bits because
a computer or Turing machine is defined to operate on such sequences and not on
every day objects such as books, pieces of fruit, molecules or sub-atomic
particles, or indeed concepts and ideas.

This last point is important. It is not possible to compute with
objects in the world, be they concrete or abstract, it is only
possible to compute with symbols representing these
objects. Another kind of relation is required for this,
$\mathbb{R} \subseteq \mathbb{S} \times \mathbb{W}$ where $\mathbb{W}$ is
the set of objects in the world. It is not possible to write down
such relations between symbols and real-world objects any more than it
is possible to write down an apple. So we have two kinds of relations
to work with: annotations which are relations among symbols 
in $\mathbb{S} \times \mathbb{S}$ and representations which map
between symbols and the world, $\mathbb{S} \times \mathbb{W}$.

Some observations are in order. First, the representation relation has
an inverse, $\mathbb{W} \times \mathbb{S}$. This is trivial and is simply ``has
the representation'' as opposed to ``represents''. Second, of course,
symbols are themselves objects in the world, so $\mathbb{S} \subset
\mathbb{W}$. Finally, relations among symbols---annotations---are likewise
objects in the world, so $\mathbb{S}\times\mathbb{S} \subset
\mathbb{W}$ also. This is useful because it means that it is possible
to represent annotations with symbols and from there articulate
relationships among them using more annotations, constructing a 
hierarchy of annotation as formalised by Buneman et al. \cite{buneman2013annotations}. 
We run into trouble
though if we try to say that representations are in the world because
$\mathbb{S}\times\mathbb{W}$ is larger than $\mathbb{W}$, and this is why
they cannot be written down. Symbols represent, annotations are
relations among symbols, and the character of representation is
fundamentally different from that of annotation.

We have enough background to explain the intuition behind the folk theory of
annotation, that there is a difference of kind between the annotation
and its object. This difference is just the same as considering a
notional pair $(x \in \mathbb{S}, -)$ \emph{qua} annotation or
\emph{qua} representation, that is, deciding the set from which the
second element of the tuple should be drawn. A similar choice is
available, \emph{mutatis mutandis}, for the inverse, $(-,
x\in\mathbb{S})$. If the unspecified element is
in $\mathbb{W} \backslash \mathbb{S}$ (i.e., tthose objects in the world that
are not symbols), there is only one choice: the relation can only
be treated as representation. If it is in $\mathbb{W} \cap \mathbb{S}$
then either interpretation is possible, and one or the other might be
more appropriate depending on the purpose or question at hand.

The ability to make this choice is no more than the ability to select an
appropriate \emph{abstraction}. Selecting an abstraction means
deciding to interpret a relation as representation and not
annotation. This is best illustrated with an example. Here is a
(representation of an) agent or substance:
\begin{align*}
  & \tikz[baseline] {
    \draw (-0.5,-0.5) rectangle (0.5,0.5);
    \draw[fill=black] (0,0.5) circle (0.1);
    \draw[fill=black] (-0.5,0) circle (0.1);
    \draw[fill=black] (0.5,0) circle (0.1);
    \node at (-0.8,0) { $u$ };
    \node at (0.8,0) { $d$ };
    \node at (0,0.8) { $b$ };
    \node at (0,0) { $A$ };
  }
  &&\mathtt{A(u,d,b)}
\end{align*}
Perhaps it is a fragment of DNA which can be connected up-stream and
down-stream to other such fragments, and it has a binding site where
RNA polymerase can attach as part of the transcription process. Some
annotations involving $A$ might be,
\begin{align*}
  (A, \mathrm{``Promoter"}) &\in \mathbb{L} \\
  (A, \mathtt{TTGATCCCTCTT}) &\in \mathbb{M}
\end{align*}
where the first is from the set of labellings, $\mathbb{L}$, and the
second is from the set of correspondences with symbols representing
nucleotide sequences, which we will call $\mathbb{M}$. A more
conventional way of writing these correspondences more closely to the
Semantic Web practice is,
\begin{center}
  \begin{minipage}{0.9\linewidth}
\begin{verbatim}
A label "Promoter"
A has sequence TTGATCCCTCTT
\end{verbatim}
  \end{minipage}
\end{center}
The labelling annotation is easy to understand. It simply provides a
friendly string for humans.

The second annotation is more challenging. It says that the DNA
fragment represented by $A$ corresponds to a certain sequence of nucleotides.
On the one hand the symbol for that sequence could simply be
taken as-is, if it does not play an explicit role in the computer
simulation of whatever interactions $A$ is involved in. That
corresponds to treating the symbol \texttt{TTGATCCCTCTT} as a
representation. It is the end of the chain; there only remains the
relation from that symbol to something in the world, which is not
something that we can write down or compute with.

On the other hand, it is equally possible to write down an annotation
on the sequence symbol that specifies the list of (symbols representing)
the nucleotides that it consists of,
\begin{center}
  \begin{minipage}{0.9\linewidth}
\begin{verbatim}
TTGATCCCTCTT consists [T,T,G,A,T,C,C,C,T,C,T,T] .
\end{verbatim}
  \end{minipage}
\end{center}
Such a verbose formulation might be useful if one had, for example, a
machine for synthesizing DNA molecules directly to implement an
experiment \emph{in vitro} for a genetic circuit that had already been
developed and tested by simulation \emph{in silico}, or a computer simulation
that worked at a very detailed level. In this case the
symbols, \texttt{A}, \texttt{C}, \texttt{T} and \texttt{G} play the role
of representing real-world objects and the symbol
\texttt{TTGATCCCTCTT} is merely a reference that can be used to find
the (list-structured) relations among them. By making this choice,
the selected abstraction has become more granular.

Another example, pertinent because while we do not yet have machines
for arbitrarily assembling DNA molecules from individuals, we do have
tools for drawing contact map diagrams, is a
rule involving this agent. This agent has a binding site which may be
occupied by an RNA-polymerase molecule at a certain rate. This could
be expressed as,
\begin{center}
\begin{minipage}{0.9\textwidth}
\begin{verbatim}
#^ r1 label "Binding of RNAp to A"
'r1' A(b!_), RNAP(s!_) -> A(b!1), RNAP(s!1) @k
\end{verbatim}
\end{minipage}
\end{center}
where now we have introduced a little bit more of the syntax that will
be more fully elaborated later for annotating rules written in a file
using the Kappa language. Here a rule is simply given a useful
human-readable label, the canonical example of annotating
something. On its own, it is useful. Imagine a summary of the contents
of a set of such rules using labels like this. For that purpose the
symbol \texttt{r1} can be considered just to represent the rule
without looking any deeper.

\begin{wrapfigure}[3]{r}{0.3\linewidth}
  \vspace{-0.9cm}
  \begin{center}
    \begin{tikzpicture}
      \begin{scope}[shift={(-1,0)}]
        \draw (-0.5,-0.5) rectangle (0.5,0.5);
        \draw (0,0) rectangle (0.5,0.5);
        \coordinate (b) at (0.5,0.25);
        \node at (0,-0.25) { \texttt{A} };
        \node at (0.25, 0.25) { \texttt{b} };
      \end{scope}
      \begin{scope}[shift={(1,0)}]
        \draw (-0.5,-0.5) rectangle (0.5,0.5);
        \draw (0,0) rectangle (-0.5,0.5);
        \coordinate (d) at (-0.5,0.25);
        \node at (0,-0.25) { \texttt{RNAp} };
        \node at (-0.25, 0.25) { \texttt{s} };
      \end{scope}
      \draw (b) -- (d);
      \draw[fill=black] (b) circle (0.05);
      \draw[fill=black] (d) circle (0.05);
      \node at (0,0.5) { \texttt{r1} };
    \end{tikzpicture}
  \end{center}
\end{wrapfigure}
For a contact map diagram, more information is needed. At right is
the diagram that corresponds to the example rule. It shows that
\texttt{A} and \texttt{RNAp} interact, that it happens through the
action of the rule \texttt{r1} and in particular involves the sites
\texttt{b} and \texttt{s}. Perhaps including which sites are involved
in the interaction is too granular and it might be desireable in some
circumstances to have a similar diagram involving just the agents and
the rules. Or perhaps more information is desired to be presented in
the diagram such as whether the rule involves creation or annihilation
of a bond, say using arrows or a broken edge. No matter the level of
granularity required, it is clear that the necessary information is
contained within the rule itself, so simply considering the symbol
\texttt{r1} to opaquely represent to the rule as an object is not
enough. Such a level of abstraction would be too coarse, it
must be elaborated further. Instead it should be considered to
represent annotations that themselves represent the structure of the
rule.

This discussion illustrates the idea of a contact map and how it can be
generated from annotations, but to elaborate the rule sufficiently to support
the production of such a diagram in practice involves a much greater amount of
annotation structure  than we 
have seen so far. A rule has a left and a right side. Each of those
has zero or more agent \emph{patterns}. A rule does not involve
agents as such, rather it involves patterns that can match
configurations of agents, so patterns then relate, \emph{intra alia},
to agents and sites, and finally bonds between sites that are either
to be matched (on the left-hand side) or created or annihilated (on
the right-hand side). It involves some work to represent a rule as annotation in
sufficient detail, but it is straightforward to do within the framework that we
have given.

\section{Annotation of Rule-Based Models}

We focus our attention on annotating models written using either the Kappa or BioNetGen language. Software tools compatible with these modeling languages are available at the following URLs:
\begin{enumerate}
\item \url{https://kappalanguage.org}
\item \url{https://github.com/RuleWorld}
\end{enumerate}


\subsection{Rationale for recommended annotation conventions}

Following our general discussion above about annotations and rule-based
models, here we move to the more technical aspects
(focusing on two languages, Kappa and BNGL) and follow the terminology and the definitions provided in Ref. \cite{Misirli2015annottation}.

Biological entities are represented by agents in Kappa and molecule
types in BNGL (we use `agent' to generically refer to both
types). Agents may include any number of sites that represent the
points of interactions between agents. For example, the DNA binding domain of a transcription factor (TF) agent can be connected to a TF
binding site of a DNA agent. Moreover, sites can have states. For
instance, a TF may have a site for phosphorylation and DNA
binding may be constrained to occur only when the state of this site
is phosphorylated.

For an agent with two sites, of which one with two internal states and the other with three, the number of possible combinations is six
(Figure~\ref{fig:agentexample}A, B). A pattern is an (possibly
incomplete) expression of an agent in terms of its internal states and
binding states. Rules specifying biological interactions consist
of patterns on the left-hand side which, when matched, produce the
result on the right-hand side
(Figure~\ref{fig:agentexample}C). Specific patterns of interest can be
declared as an observable of a model (i.e., a simulation output).

\begin{figure} [thp]
\lstsetrule
\begin{lstlisting}
$\textbf{A}: An agent definition$
A(site1~u~v, site2~x~y~z)

$\textbf{B}: Possible combinations of internal states$
A(site1~u,site2~x)
A(site1~u,site2~y)
A(site1~u,site2~z)
A(site1~v,site2~x)
A(site1~v,site2~y)
A(site1~v,site2~z)

$\textbf{C}: An example binding rule$
A(site1~v,site2~z),A(site1~v,site2~y) 
    -> A(site1~v!1,site2~z),A(site1~v!1,site2~y) @kf
\end{lstlisting}
  \caption{\textbf{A.} An agent with two sites. \texttt{site1} has two
    possible internal states while \texttt{site2} has
    three. \textbf{B.} This agent can be used in six different ways
    depending on the internal states of its sites. \textbf{C.} A rule
    that specifies how agent \texttt{A} forms a dimer when the state
    of \texttt{site1} is \texttt{v} and the states of \texttt{site2}
    are \texttt{z} and \texttt{y}, respectively. The symbol
    \texttt{!n} means that the sites where it appears are bound (connected) together. The constant \texttt{kf} denotes the kinetic rate
    associated with the rule.
  }
  \label{fig:agentexample}
\end{figure}

It is important to highlight that while the syntactic definition of an
agent identifies sites and states in rule-based models, the semantics
of sites and states is usually clear only to the modeller. Cleary, if one wishes to have machine access, then this information must be exposed in a structured way. The key idea of the approach presented in Ref. \cite{Misirli2015annottation} and that we review in what follows, is to extend the syntax of rule-based models to incorporate annotations.

Existing metadata resources include machine readable controlled
vocabularies and ontologies and Web services providing standard access to
external identifiers and guidelines for the use of these
resources. For example, the Minimum Information Requested in the
Annotation of Models (MIRIAM) standard~\cite{Novere2005} provides a standard for the minimal information required for the
annotation of models.

Following Ref. \cite{Misirli2015annottation} we suggest that entities in
models should be linked to external information through the use of unique and unambiguous Uniform Resource Identifiers (URIs), which are embedded within models. The uniqueness and
global scope of these URIs are then crucial for
\textit{disambiguation} of model agents, variables and rules.

We also choose to represent annotations using the Resource Description Framework
(RDF) data model~\cite{rdf_concepts,rdf_xml_syntax} 
as statements or binary predicates. A statement can link
a modelling entity to a value using a standard qualifier term
(predicate), which represents the relationship between the entity and
the value. These qualifiers often come from controlled vocabularies or
ontologies in order to unambiguously identify the meaning of modelling
entities. URIs are used as values to link these entities to external
resources, and hence to a large amount of biological information by keeping
the number of annotations minimal. The links themselves are typed,
again with URIs. The qualifiers and resources to which they refer  are drawn from ontologies that encode the Description
Logic~\cite{w3c_owl_features} for a particular domain.

Semantics can be unified by means of metadata with controlled
vocabularies. There are several metadata standard initiatives that
provide controlled vocabularies from which standard terms can be taken. For instance, metadata terms provided by the Dublin Core
Metadata Initiative (DCMI)~\cite{dcmi} or BioModels qualifiers can be
used to describe modelling and biological
concepts~\cite{Novere2005b,Li2010}. On the other hand, ontologies such as the Relation
Ontology provide formal definitions of relationships that can be used
to describe modelling entities~\cite{Smith2005}. There are also
several other ontologies and resources that are widely used to
classify biological entities represented in models with standard
values~\cite{Swainston2009}: the Systems Biology Ontology
(SBO)~\cite{Courtot2011} to describe types of rate parameters; the
Gene Ontology (GO)~\cite{GOC2001} and the Enzyme Commission (EC)
numbers~\cite{Bairoch2000} to describe biochemical reactions; the
Sequence Ontology (SO)~\cite{Eilbeck2005} to annotate genomic features
and unify the semantics of sequence annotation; the BioPAX
ontology~\cite{Demir2010} to specify types of biological molecules and
the Chemical Entities of Biological Interest
(ChEBI)~\cite{Degtyarenko01012008} terms to classify chemicals. URIs
of entries from biological databases, such as
UniProt~\cite{Magrane2011} for proteins and KEGG~\cite{Kanehisa2008}
for reactions, can also be used to uniquely identify modelling
entities.

Access to data should be unified and
this can be done by accessing external resources through
URIs using MIRIAM or Identifiers.org URIs~\cite{Juty01012012}. It should be
noted that MIRIAM
identifiers are not resolvable directly over the Internet and require out of
band knowledge to retrieve additional information though they are unique and unambiguous. These URIs consist of
collections and their terms, which may represent external resources
and their entries respectively. For example, the MIRIAM URI
\url{urn:miriam:uniprot:P69905} (\textit{see} \textbf{Note 1}) 
and the Identifiers.org URI
\url{http://identifiers.org/uniprot/P69905} can be used to link
entities to the P69905 entry from UniProt.  The relationships between
modelling entities, annotation qualifiers and values can be
represented using RDF graphs.

We recommend to use RDF syntax that represents
knowledge as $(\mathrm{subject}$, $\mathrm{predicate}$,
$\mathrm{value})$ triples, in which the subject can be an anonymous
reference or a URI, the predicate is a URI and the object can be a
literal value, an anonymous reference or a URI. 

Subjects and objects may refer to an ontology term, an external resource or an entity
within a model. RDF graphs can be then serialized in different
formats such as XML or the more human readable Turtle
format \cite{rdf_turtle}. Modelling languages such
as the Systems Biology Markup Language (SBML)~\cite{Hucka2003},
CellML~\cite{Cuellar2003,Hedley2001} and Virtual Cell Markup
Language~\cite{Moraru2008} are all XML-based and provide facilities to
embed RDF/XML annotations~\cite{Endler2009}.

Moreover, there are also other exchange languages,
such as BioPAX and the Synthetic Biology
Open Language (SBOL)~\cite{Galdzicki2012,Galdzicki2014}, that can be
serialised directly as RDF/XML allowing custom annotations to be embedded.

Following the suggestion of Misirli et al. \cite{Misirli2015annottation} one can
extend the use of RDF and MIRIAM annotations to describe a syntax to
store machine-readable annotations and an ontology to facilitate the
mapping between rule-based model entities and their annotations.
We illustrate annotations using terms from this ontology and propose some examples.

\subsection{Conventions for annotating Kappa- and BNGL-formatted models}

Here, we review the syntax originally defined by Misirli et al. \cite{Misirli2015annottation} for storing annotations.
We start by noticing that a common approach, when trying to add
additional structured information to a language where it is
undesirable to change the language itself, is to define a special way
of using comments. This practice is established for structured
documentation or ``docstrings'' in programming
languages~\cite{acuff1988ksl,stallman1992gnu}. The idea is to use this same approach so that models written using the
conventions that we describe here do not require modification of
modelling software, such as KaSim~\cite{kasim} or RuleBender~\cite{Xu2011}.

For this reason, we  use the language's comment delimiter followed by the
`\texttt{\textasciicircum{}}' character to denote annotations in the
textual representation of rule-based languages. Kappa and BNGL both
use the `\texttt{\#}' symbol to identify comment lines, so in the case
of these languages, comments containing annotations are signalled by a
line beginning with `\texttt{\#\textasciicircum{}}'. This
distinguishes between comments containing machine-readable annotations and comments
intended for direct human consumption. Annotation data for a single modelling
entity or a model itself can be declared over several lines and each
line is prefixed with the `\texttt{\#\textasciicircum{}}' symbol.

Annotations are then serialised in the RDF/Turtle format. We claim that this leads
to a good balance between the need for a machine-readable syntax and a
human readable textual representation. Rule-based modelling
languages are themselves structured text formats designed for this
same balance, so RDF/Turtle is more suitable than the XML-based
representations of RDF. 

Annotations for a single rule-based model entity are a list of
statements. It is important to stress that annotations may refer to other annotations within the same
model. When all the lines corresponding to a rule-based model and the
annotation delimiter symbols are removed, the remaining RDF lines
can represent a single RDF document. This enables annotations to be
quickly and easily extracted without special tools (\textit{see} \textbf{Note 2}).

In textual rule-based models, it is difficult to store annotations
within a modelling entity since Kappa and BNGL represent
modelling entities such as agents and rules as single lines of
text. As a result, there is no straightforward location to attach annotations
to an entity.  Following Ref. \cite{Misirli2015annottation} we achieve the
mapping between a modelling entity and its annotations by defining an
algorithm to construct a URI from the symbol used in the modelling
language. The algorithm generates unique and unambiguous prefixed
names that are intended to be interpreted as part of a Turtle
document. The algorithm simply constructs the local part of a prefixed name
by joining symbolic names in the modelling language with the `:'
character, and prepending the empty prefix, `:'. This means that one
must satisfy the condition that the empty prefix is defined for this
use. Using this algorithm, we can derive a globally unique reference for the \texttt{y}
internal state of site \texttt{site2} of agent \texttt{A} from
\texttt{A(site1\textasciitilde{}u\textasciitilde{}v,site2\textasciitilde{}x\textasciitilde{}y\textasciitilde{}z)}
as \texttt{:A:site2:y}. 

In Kappa, rules do not have symbolic names but each rule can be
preceded by free text surrounded by single quotes.  We require this free text to be consistent with the local name syntax in the Turtle and
SPARQL~\cite{rdf_sparql} languages. If this requirement is satisfied, identifiers for
subrules are created by just adding their position index, based on one, to
the identifier for a rule (see Figure \ref{fig:ruleannotation}B).
A similar restriction is placed on other tokens used in the models;
agent and site names, variable and observable names must all conform
to the local name syntax.
 
Controlled vocabularies such as BioModels.net qualifiers are formed of
\textit{model} and \textit{biology} qualifiers. The former offers
terms to describe models. BioModels.net qualifiers are also
appropriate to annotate rule-base models, but additional qualifiers
are needed to fully describe rule-based models. These are specific to
the annotation of rule-based models and this is done by using a
distinct ontology -- the \emph{Rule-Based Model Ontology} -- in the
namespace \url{http://purl.org/rbm/rbmo#} conventionally abbreviated
as \texttt{rbmo} (we omit the prefix if there is no risk of
ambiguity). Each qualifier is constructed by combining this namespace
with an annotation term. A subset of significant terms are listed in
Table~1 while the full
ontology is available online at the namespace URI.

In the \texttt{rbmo} vocabulary, the \rbmterm{Model} classes such as \rbmterm{Kappa} and
\rbmterm{BioNetGen} specify the type of the model being annotated.
The term \rbmterm{Agent} is used to declare physical molecules. Hence,
the \rbmterm{Agent} class can represent agents and tokens in Kappa, or
molecule types in BioNetGen. \rbmterm{Site} and \rbmterm{State}
represent sites and states in these declarations respectively. Rules
are identified using \rbmterm{Rule}.
The predicates \rbmterm{hasSite} and \rbmterm{hasState} and their
inverses are used to annotate the links between agents, sites and
internal states declarations. Table~1 reviews the terms
related to the declaration of the basic entities from which models are
constructed. We assume that the terms that start with an uppercase
letter are types (In the sense of \rbmterm{rdf:type}, and also in this
instance \rbmterm{owl:Class}) for the entities in the model which the
modeller could be expected to explicitly annotate. The predicates
begin with a lowercase letter and are used to link entities to their
annotations.

Table~2 includes terms to facilitate
representation of rules in RDF. This change of representation
(materialization), from Kappa or BNGL to RDF is something that
can easily be automated and a tool is already available (for models
written in Kappa).

This representation in RDF is helpful for analysis of models because it 
merges the model itself with the metadata in a uniform way easy to
query.  Annotations that cannot be derived from the model
(as well as the model itself) are written explicitly in RDF/Turtle
using the terms from Table~1 embedded in comments using
a special delimiter. Extra statements can then be derived by
parsing and analyzing the model using terms from
Table~2 and the same naming convention from the
algorithm previously described.  These statements are then merged with
the externally supplied annotations to obtain a complete and uniform
representation of all the information about the model.

The open-ended nature of the RDF data model means that it is possible
to freely incorporate terms from other ontologies and vocabularies,
including application-specific ones.  In this respect, two terms are
crucial. The \rbmterm{dct:isPartOf} predicate from DCMI Metadata Terms
is used to denote that a rule or agent declaration \emph{is part of} a
particular model (or similarly with its inverse,
\rbmterm{dct:hasPart}).

The \rbmterm{bqiol:is} predicate from the \emph{Biomodels.net Biology
  Qualifiers} is used to link internal states of sites to indicate
their biological meaning. This term is chosen because it denotes a
kind of identification that is much weaker than the logical
replacement semantics of \rbmterm{owl:sameAs}. Using the latter would
imply that everything that can be said about the site \emph{qua}
biological entity can also be said about the site \emph{qua} modelling
entity. Clearly, these are not the same and identifying them
in a strong sense would risk incorrect results when computing with the
annotations. 

Table~3 enumerates useful ontologies and
vocabularies with their conventional prefixes to annotate rule-based
models. This list is not exhaustive and can be extended.

\subsection{Adding annotations to model-definition files}

Here, we demonstrate how the suggested annotations can be
added to rule-based models. Again we follow the methodology originally presented in  Ref.~\cite{Misirli2015annottation}. 

\begin{figure} [ht]
\lstsetrule
\begin{lstlisting}
#^@prefix : <http://.../tcs.kappa#>.
#^@prefix rbmo: <http://purl.org/rbm/rbmo#>.
# ... other prefixes elided ...
#^@prefix dct: <http://purl.org/dc/terms/>.
#^@prefix foaf: <http://xmlns.com/foaf/0.1/>.

#^ :kappa a rbmo:Kappa  ;
#^   dct:title "TCS_PA Kappa model" ;
#^   dct:description
#^     "Two component systems and promoter architectures" ;
#^   dct:creator "Goksel Misirli", "Matteo Cavaliere"; 
#^   foaf:isPrimaryTopicOf <https://.../tcs.kappa> .
\end{lstlisting}
  \caption{An example model annotation (as in~\cite{Misirli2015annottation}), with details about its name, description, creators and online repository location. The prefix definitions required to annotate the model are defined first, and the empty prefix is defined for the model namespace itself.}
  \label{fig:modelannotation}
\end{figure}

\begin{figure}
\lstsetrule
\begin{lstlisting}
$\textbf{A}:$
#^:ATP a rbmo:Agent ;
#^  bqbiol:isVersionOf chebi:CHEBI:15422 ;
#^  biopax:physicalEntity biopax:SmallMolecule .
%token: ATP()
$\textbf{B}:$
#^:Kinase a rbmo:Agent ;
#^  rbmo:hasSite :Kinase:psite ;
#^  bqbiol:is uniprot:P16497 ;
#^  biopax:physicalEntity biopax:Protein ;
#^  ro:hasFunction go:GO:0000155 .
#^:Kinase:psite a rbmo:Site ;
#^  rbmo:hasState :Kinase:psite:u, :Kinase:psite:p .
#^:Kinase:psite:u a rbmo:State ;
#^  bqiol:is pr:PR:000026291 .
#^:Kinase:psite:p a rbmo:State ;
#^  bqiol:is psimod:MOD:00696 .
%agent: Kinase(psite~p~u)
$\textbf{C}:$
#^:pSpo0A a rbmo:Agent ; 
#^  rbmo:hasSite :pSpo0A:tfbs ;
#^  bqbiol:isVersionOf so:SO:0000167 ;
#^  biopax:physicalEntity biopax:DnaRegion ;
#^  sbol:nucleotides "ATTTTTTTAGAGGGTATATAGCGGTTTTGTCGAATGTAAACATGTAG" ;
#^  sbol:annotation :pSpo0A_annotation_28_34 .	
#^:pSpo0A:tfbs a rbmo:Site ;
#^  bqbiol:isVersionOf so:SO:0000057 ;
#^  biopax:physicalEntity biopax:DnaRegion ;
#^  sbol:nucleotides "TGTCGAA" .
#^:pSpo0A_annotation_28_34 a sbol:SequenceAnnotation ;
#^  sbol:bioStart 28;
#^  sbol:bioEnd 34 ;
#^  sbol:subComponent :pSpo0A:tfbs .
%agent: pSpo0A(tfbs)
$\textbf{D}:$
#^:Spo0A a rbmo:Agent .
%agent: Spo0A(psite~p~u)
#^:Spo0A_p a rbmo:Observable ;
#^   ro:has_function go:GO:0045893 .
%obs: 'Spo0A_p' Spo0A(psite~p)
\end{lstlisting}
 \caption{Examples of agent annotations for \textbf{A.} An ATP token
   agent. \textbf{B.} A kinase agent with phosphorylated and
   unphosphorylated site. \textbf{C.} A promoter agent with a TF
   binding site. \textbf{D.} An agent and an associated observable for
   the phosphorylated Spo0A protein, which can act as a TF.}
  \label{fig:agentannotation}
\end{figure}

Annotations are added by simply adding a list of prefix definitions representing
annotation resources providing relevant terms for the annotation of
all model entities (such as agents and rules). These definitions are
followed by statements about the title and description of the model,
using the \rbmterm{title} and \rbmterm{description} terms from
\emph{Dublin Core}.  Annotations can be expanded to include model
type, creator, creation time, and its link to an entry in a model database
(Figure~\ref{fig:modelannotation}).

Table~4 shows how distinct entities in a
model can be annotated using terms from \texttt{rbmo} and from other
vocabularies. Figure~\ref{fig:agentannotation} shows examples of
Agent annotations. In Figure~\ref{fig:agentannotation}A the ATP token
is annotated as a small molecule with the identifier 15422 from
ChEBI. Agents without sites can also be annotated in a similar way.
In Figure~\ref{fig:agentannotation}B, the agent is specified to be a
protein using the \rbmterm{biopax:Protein} value for the
\rbmterm{biopax:physicalEntity} term. This protein agent is annotated
as P16497 from UniProt, which is a protein kinase (i.e., an enzyme that
phosphorylates proteins) involved in the process of sporulation.
It has a site with the phosphorylated and unmodified states,
which are annotated with corresponding terms from
the Protein Modification Ontology~\cite{Montecchi-Palazzi2008}.

The \rbmterm{ro:hasFunction} term associates the agent with the GO's
histidine kinase molecular function term \rbmterm{GO:0000155}. In
Figure~\ref{fig:agentannotation}C, a promoter agent with a TF binding
site is represented. Both the promoter and the operator agents are of
``DnaRegion'' type, and are identified with the \rbmterm{SO:0000167}
and \rbmterm{SO:0000057} terms. Although the nucleotide information
can be linked to existing repositories using the \rbmterm{bqbiol:is}
term, for synthetic sequences agents can directly be annotated using
SBOL terms. The term \rbmterm{sbol:nucleotides} is used to store the
nucleotide sequences for these agents.  A parent-child relationship
between the promoter and the operator agents can be represented using
an \rbmterm{sbol:SequenceAnnotation} RDF resource, which allows the
location of an operator subpart to be specified.

This approach can be used to annotate a pattern with a specific entry
from a database (patterns can also be stated as observables of the
model). For instance, Figure~\ref{fig:agentannotation}D shows an
example of such an observable. \texttt{Spo0A\_p} represents the
phosphorylated protein, which acts as a TF and is defined as an
observable.

Figure~\ref{fig:ruleannotation} demonstrates annotation of rules. The
first rule (Figure~\ref{fig:ruleannotation}A) describes the binding of the
LacI TF to a promoter. This biological activity is described using the
\rbmterm{GO:0008134} (\textit{transcription factor binding}) term. In
the second example (Figure~\ref{fig:ruleannotation}B), a
phosphorylation rule is annotated. The rule contains a subrule
representing ATP to ADP conversion. This subrule is linked to the
parent rule with the \rbmterm{hasSubrule} qualifier. Moreover, the annotation of the rate for this rule is presented in
Figure~\ref{fig:ruleannotation}C.
The annotated Kappa and BNGL models for a two-component system
(TCS), controlling a simple promoter architecture can be found online (\textit{see} \textbf{Note 3}).

Finally, in Figure~\ref{fig:materialised} we present the fragment of a
specific rule (taken from the TCS Kappa model) materialised using the
\texttt{krdf} tool. The tool generates a version of the rules
themselves in RDF together with the annotations (in this way the
entire model is presented in a more uniform way).

\begin{figure} [ht]
\lstsetrule
\begin{lstlisting}
$\textbf{A}:$
#^:LacI.pLac a rbmo:Rule	;
#^  bqbiol:isVersionOf go:GO:0008134 ;
#^  dct:title "Dna binding" ;
#^  dct:description "TF1 binds to the promoter" .	
'LacI.pLac' Target(x~p), Promoter(tfbs1,tfbs2) <-> Target(x~p!1), Promoter(tfbs1!1,tfbs2) @kf,kr
$\textbf{B}:$
#^:S_phosphorylation a rbmo:Rule ;
#^   bqbiol:isVersionOf sbo:SBO:0000216 ;
#^   dct:title "S Phosphorylation" ;
#^   dct:description "S is phosphorylated" ;
#^   rbmo:hasSubrule :S_phosphorylation:1 .	
#^:S_phosphorylation:1 a rbmo:Rule ;
#^   bqbiol:isVersionOf sbo:SBO:0000216 ;
#^   dct:title "ATP -> ADP" ;
#^   dct:description "ATP to ADP conversion" .
'S_phosphorylation' S(x~u!1), K(y!1) | 0.1:ATP -> S(x~p), K(y) | 0.1:ADP  @kp
$\textbf{C}:$
#^:kp a sbo:SBO:0000002 ; 
#^  bqbiol:isVersionOf sbo:SBO:0000067 ;
#^  dct:title "Phosphorylation rate" .
\end{lstlisting}
  \caption{Annotating rules and variables. \textbf{A.} TF DNA binding rule. \textbf{B.} Phosphorylation rule with a subrule for the ATP to ADP conversion. \textbf{C.} Annotation of a phosphorylation rate variable.}
  \label{fig:ruleannotation}
\end{figure}

\begin{figure}[ht]
\lstsetrule
\begin{lstlisting}

:As1As2Spo0A_to_As2Spo0A a rbmo:Rule ;    
  dct:title "Cooperative unbinding" ;
  rbmo:lhs [
    a rbmo:Pattern ;
    rbmo:agent :Spo0A ;
    rbmo:status [
      rbmo:isBoundBy :As1As2Spo0A_to_As2Spo0A:left:1 ;
      rbmo:isStatusOf :Spo0A:DNAb ;	
      a rbmo:BoundState ;	
    ], [
      rbmo:internalState :Spo0A:RR:p ;
      rbmo:isStatusOf :Spo0A:RR ;
      a rbmo:UnboundState ;	
    ] ;
  ].
\end{lstlisting}
  \caption{Fragment of the RDF representation of a materialised rule
    obtained by merging the metadata supplied by the model author with
    an RDF representation of the rule.  The left hand side of the rule
    contains a pattern involving \rbmterm{:Spo0A} and that there are
    two pieces of state information: The first one refers to the
    \rbmterm{:Spo0A:DNAb} site, and it is bound to something (that can
    only be recovered using the rest of the model, not presented
    here). The second refers to the \rbmterm{:Spo0A:RR} site, it has a
    particular internal state, and it is unbound.}
  \label{fig:materialised}
\end{figure}

\subsection{How to use annotations}

The framework we have described can be coupled to the development of tools
that allow one to extract and analyze the annotations embedded in a model.
Several tools are currently under development.  We demonstrate here
the \texttt{krdf} tool that can be used for checking duplication of
rules and inconsistencies between different parts of a model, basic
problems encountered when composing and creating biological
models~\cite{Blinov2008, Lister2009}. Another application is to draw
an annotated contact map visualising the entities involved, the
interactions and the biological information stored in the annotations
-- this merges the classical notion of contact map used to illustrate Kappa and BNGL
models~\cite{Danos2004, Danos2009} with biological semantics.

The \texttt{krdf} tool operates on Kappa models and has several modes
of operation that can provide increasingly more information about a
model. The first, selected with the \texttt{-a} option, extracts the
modeller's annotations. The second mode, selected with the \texttt{-m}
option, \emph{materialises} the information in the rules themselves
into the RDF representation (as illustrated in
Figure~\ref{fig:materialised}). Finally the \texttt{-n} option
\emph{normalises} the patterns present in the rules according to their
declarations.

Once a complete uniform representation of the model in RDF has been
generated, one can query it using SPARQL with a tool such as
\texttt{roqet}~\cite{beckett2015redland}. For example, a SPARQL
query can deduce a contact map -- pairings of sites in agents that
undergo binding and unbinding according to the rules in a
model. These pairings form a graph that can be visualised using tools
such as GraphViz~\cite{ellson2002graphviz}. With an appropriate
query (\textit{see} \textbf{Note 4}), \texttt{roqet} can output the result in a GraphViz-compatible
format. A more sophisticated manipulation (\textit{see} \textbf{Note 5}) can extract
annotations from the RDF representation of the TCS example model and
easily create a richly annotated contact map diagram (Figure
~\ref{fig:contact}). In this way, biological information extracted
from the annotations can be added to the agents, sites and
interactions (using GraphViz for rendering) (\textit{see} \textbf{Note 6}).

\begin{figure}[ht]
  \begin{center}
    \includegraphics[scale=0.5]{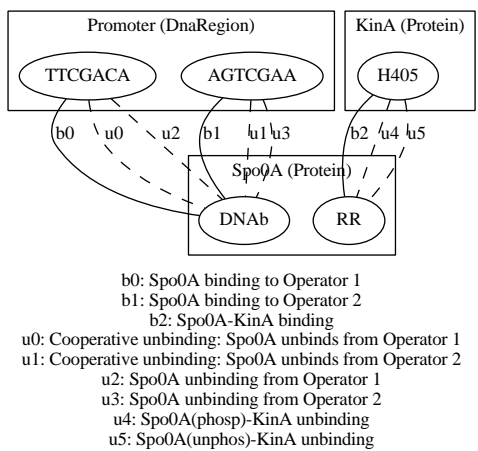}
    \caption{Contact map generated by a SPARQL query on the RDF
      materialisation of the TCS example in Kappa. Biological
      information concerning the agents, rules and sites, types of the
      molecules, DNA sequences and typology of the interaction, are
      extracted automatically from the model annotations.
      This figure is a reproduction of Fig. 6 in Ref. \cite{Misirli2015annottation};
      no changes have been made. The figure is used under the terms of the CC-BY
      license (\texttt{https://opendefinition.org/licenses/cc-by/}).}
    \label{fig:contact}
  \end{center}
\end{figure}
\medskip

Moreover, one can easily create a query that implements a join operation
on the property of \rbmterm{bqbiol:is}, enforcing a stronger form of
identity semantics than this predicate is usually given. A filter
clause is necessary to prevent a comparison of a rule with itself (see
the SPARQL query in Figure~\ref{fig:duplicate}). In this way, the discussed annotations could also be used to detect duplication of rules
(e.g., obtained when combining different biological models).

\FloatBarrier
\begin{figure}[!h]
\lstsetrule
\begin{lstlisting}
SELECT DISTINCT ?modelA ?ruleA ?modelB ?ruleB
WHERE {
  ?ruleA a rbmo:Rule;
      dct:isPartOf ?modelA;
      bqbiol:is ?ident.
  ?ruleB a rbmo:Rule;
      dct:isPartOf ?modelB;
      bqbiol:is ?ident.
  FILTER (?ruleA != ?ruleB)
}
\end{lstlisting}
  \caption{Detection of duplicate rules.}
  \label{fig:duplicate}
\end{figure}
\FloatBarrier
\medskip

Another possible application of the presented annotation schema is the checking of 
inconsistencies in a rule-based model.  This can be done in several different ways. A simple way is to use
the replacement semantics of \rbmterm{owl:sameAs}.  A statement of the
form \rbmterm{a owl:sameAs b} means that every statement about
\rbmterm{a} is also true if \rbmterm{a} is replaced by \rbmterm{b}. In
particular if we have statements about the types of \rbmterm{a} and
\rbmterm{b}, and these types are disjoint, the collection of
statements is unsatisfiable (hence, the model has been found to be
inconsistent). Then, an OWL reasoner such as
HermiT~\cite{shearer2008hermit} or Pellet~\cite{sirin2007pellet} can
derive that \rbmterm{a} and \rbmterm{b} have type
\rbmterm{owl:Nothing}.

This can be implemented with the following work-flow (here only sketched): (i) generate the
fully materialised RDF version of a model using \texttt{krdf}. For each use of \rbmterm{bqbiol:is}, add a new
statement using \rbmterm{owl:sameAs}; (ii) retrieve all ontologies
that are used from the Web. For each external vocabulary term with
\rbmterm{bqbiol:is} or \rbmterm{bqbiol:isVersionOf} retrieve a
description and any ontology that it uses (recursively). Merge all of
these into a single graph. This graph contains the complete model and
annotations, with entities linked using a strong form of equality to
external vocabulary terms, and descriptions of the meaning of these
vocabulary terms; (iii) the reasoner can be used to derive terms that
are equivalent to \rbmterm{owl:Nothing} and if any of these terms is
found then an inconsistency has been identified. Using the proof
generation facilities of OWL reasoners, the sequence of statements
required to arrive at \rbmterm{foo rdf:type owl:Nothing} can be
reproduced (in this way, the initial source of the inconsistency can be also
identified).

\subsection{Closing remarks}

In this chapter we have reviewed the recent proposal to incorporate
annotations into rule-based models, following the approach recently presented in Ref.
\cite{Misirli2015annottation}. We have also discussed in a more general way the role of annotations and how they are strongly related to the notion of abstraction. In general, for consistency, we have followed the terms originally defined in Ref. \cite{Misirli2015annottation}. However, the suggested standardized terms can
be used in a complementary manner with existing metadata resources
such as MIRIAM annotations and URIs, and existing controlled
vocabularies and ontologies.  Although, the approach has only
described the annotations of Kappa- and BNGL-formatted model-definition files, it can be
easily applied to other formats for rule-based models.

In particular, PySB~\cite{Lopez2013} already includes a list of MIRIAM annotations at
the model level, and can be extended to include the type of
annotations described here. SBML's
\texttt{multi} package (\textit{see} \textbf{Note 7}) \cite{Zhang2018}
is intended to standardise the exchange of rule-based
models.  The entities in this format inherit the annotation property
from the standard SBML and can therefore include RDF
annotations. These SBML models could thus be imported or exported by
tools such as KaSim or BioNetGen/RuleBender, avoiding the loss of any biological
information.

It is important to remark that annotations are also useful for automated conversions between
different formats. Conversion between rules and reaction networks is
already an ongoing research subject~\cite{Blinov2008}, and the
availability of annotations can play an important role for reliable
conversion and fine-tuning of models~\cite{Tapia, Harris}.  It is
straightforward to use the framework presented and automatically map
agents and rules to glyphs~\cite{Chylek2011} or to convert models into
other visual formats such as SBGN or genetic circuit
diagrams~\cite{Misirli2011}.

More generally, annotations are designed for machine readability and
can be produced computationally (e.g., by model repositories).  This
can be done by developing APIs and tools to access a set of
biological parts~\cite{Cooling, Misirli2014} that will incorporate
rule-based descriptions and will be annotated with the proposed schema. This
will open the possibility of composing (stitching together) rule-based models
extracted from distinct repositories.  Tools such as
Saint~\cite{Lister2009} and SyBIL~\cite{Blinov2010} could be extended
to automate the annotation of rule-based models. In this way, the extensive
information available in biological databases and the literature can
be integrated and made available via rule-based models, taking
advantage of the syntax and the framework presented here and elsewhere. 

One of the ultimate goals is to use annotations as a facilitator of automatic 
composition of rule-based models.  As recently suggested by Misirli et al. \cite{Misirli2016} the proposed schema can be used to automate the design of biological systems using a rule-based model with a  workflow  that combines the definition of modular templates to instantiate rules for basic biological parts. 
The templates, defining rule-based models for basic biological parts (\textit{see} \textbf{Note 8}), can be associated with quantitative parameters to create particular parts models, which can then be merged into executable models. Such models may be annotated using the reviewed schema leading to a feasible protocol to automate their composition for the scalable modelling of synthetic systems \cite{Misirli2016}.

The described annotation ontology for rule-based models can be found at
\url{http://purl.org/rbm/rbmo} while the tool and all the presented
examples can be found at \url{http://purl.org/rbm/rbmo/krdf}.

\section{Notes}

\begin{enumerate}

\item A dereferenceable URI using the MIRIAM Web service is
  \url{http://www.ebi.ac.uk/miriamws/main/rest/resolve/urn:miriam:uniprot:P69905}

\item For example, on a UNIX system, the following pipeline could be
  used:
  \begin{center}
  \texttt{grep
    \char13\textasciicircum\#\textbackslash\textasciicircum\char13 |
    sed
    \char13{}s/\textasciicircum\#\textbackslash\textasciicircum//\char13{}
  }
  \end{center}
  \vspace{-1\baselineskip}
  \vspace{0.25in}

\item The files \texttt{tcs.kappa} and \texttt{tcs.bngl}  are available in the \url{http://purl.org/rbm/rbmo/examples} directory.

\item See the \texttt{binding.sparql} file in the krdf directory.

\item See the \texttt{contact.py} script in the krdf directory.

\item The tool assumes that only single instances of an agent are involved in a rule. It can be generalized.

\item See \url{http://sbml.org/Documents/Specifications/SBML_Level_3/Packages/multi} for details.

\item These are available at \url{http://github.com/rbm/composition}.

\end{enumerate}

\section*{Acknowledgement}
The Engineering and Physical Sciences Research Council grant EP/J02175X/1 (to V.D. and M.C.), the European Union's Seventh Framework Programme for research, technological development and demonstration grant 320823 RULE (to W.W., R.H-Z, V.D.).

\bibliographystyle{spbasicunsorted}
\bibliography{annotation}

\clearpage
\section*{Tables}

\subsection*{Table 1}

\begin{table}[!ht]
{\begin{tabular}{p{0.35\hsize}p{0.58\hsize}}
\textbf{Term} & \textbf{Description} \\
\rbmterm{Kappa}, \rbmterm{BioNetGen} & Model types.\\
\rbmterm{Agent} & Type for declarations of biological entities.\\
\rbmterm{Site} & Type for sites of \rbmterm{Agent}s.\\
\rbmterm{State} & Type for internal states of \rbmterm{Site}s.\\
\rbmterm{hasSite}, \rbmterm{hasState}, \rbmterm{siteOf}, \rbmterm{stateOf} & Predicates for linking \rbmterm{Agent}s, \rbmterm{Site}s and \rbmterm{State}s. \\
\rbmterm{Rule} & Type for interactions between agents. \\
\rbmterm{hasSubrule}, \rbmterm{subruleOf} & Specifies that a rule has a subrule (i.e., KaSim subrules).\\
\rbmterm{Observable} & Type for agent patterns counted by a simulation.\\
\end{tabular}}{}
\end{table}

\clearpage

\subsection*{Table 2}

\begin{table}[!ht]
{\begin{tabular}{p{0.20\hsize}p{0.73\hsize}}
\textbf{Term} & \textbf{Description} \\
\rbmterm{Pattern} & Type of a pattern as it appears in a \rbmterm{Rule} or \rbmterm{Observable}. \\
\rbmterm{lhs}, \rbmterm{rhs} & Predicates for linking a \rbmterm{Rule} to its left and right hand side \rbmterm{Pattern}s.\\
\rbmterm{pattern} & Predicate for linking an \rbmterm{Observable} to the patterns that it matches.\\
\rbmterm{agent} & Predicate for linking a \rbmterm{Pattern} and a site within it to the corresponding \rbmterm{Agent}. \\
\rbmterm{status} & Specifies a status of a particular \rbmterm{Site} (and \rbmterm{State}) in a \rbmterm{Pattern}. \\ 
\rbmterm{isStatusOf}, \rbmterm{internalState} & Predicates for linking a status in a \rbmterm{Pattern} to corresponding \rbmterm{Site} and \rbmterm{State} declarations. \\ 
\rbmterm{isBoundBy} & Specifies the bond that a \rbmterm{Site} is bound to in a particular \rbmterm{Pattern}. Bonds are identified via URIs. \\ 
\rbmterm{BoundState}, \rbmterm{UnboundState} & Terms denoting that a \rbmterm{Site} in a \rbmterm{Pattern} is bound or unbound. \\ 
\end{tabular}}{}
\end{table}

\clearpage

\subsection*{Table 3}

\begin{table}[!ht]
{\begin{tabular}{p{0.15\hsize}p{0.77\hsize}}
\textbf{Prefix}	& \textbf{Description}  \\
\texttt{rbmo} & Rule-based modelling ontology (presented in this paper)\\ 
\texttt{dct} & Dublin Core Metadata Initiative Terms (\path{http://www.dublincore.org/documents/dcmi-terms})\\ 
\texttt{bqiol} & BioModels.net Biology Qualifiers \cite{Li2010}\\
\texttt{go} & 	Gene Ontology \cite{GOC2001}\\
\texttt{psimod} & Protein Modification Ontology \cite{Montecchi-Palazzi2008}\\
\texttt{so} & Sequence Ontology \cite{Eilbeck2005}\\
\texttt{sbo} & Systems Biology Ontology \cite{Courtot2011}\\
\texttt{chebi} & Chemical Entities of Biological Interest Ontology \cite{Degtyarenko01012008}\\
\texttt{uniprot} & UniProt Protein Database \cite{Magrane2011}\\
\texttt{pr} & Protein Ontology \cite{Natale01012011}\\
\texttt{ro} & OBO Relation Ontology \cite{Smith2005}\\
\texttt{owl} & Web Ontology Language (\path{http://www.w3.org/TR/owl-features})\\
\texttt{sbol} & The Synthetic Biology Open Language \cite{Galdzicki2012,Galdzicki2014} \\
\texttt{foaf} & Friend of a Friend Vocabulary (\path{http://xmlns.com/foaf/spec})\\
\texttt{ipr} & InterPro \cite{Mulder2008}\\
\texttt{biopax}	& Biological Pathway Exchange Ontology Ontology \cite{Demir2010} \\
\end{tabular}}{}
\end{table}

\clearpage

\subsection*{Table 4}

\begin{table}[t]
\footnotesize
{\begin{tabular}{p{4.25cm}p{10cm}} 
\textbf{Term} & \textbf{Annotation Values}\\
\hline
\underline{Agent declarations:}  \\
\rbmterm{rdf:type} & \rbmterm{Agent} \\
\rbmterm{dct:isPartOf} & Identifier for the \rbmterm{Model}.\\
\rbmterm{hasSite} & Identifier of a \rbmterm{Site}.\\ 
\rbmterm{biopax:physicalEntity} & A \rbmterm{biopax:PhysicalEntity} term, e.g. \rbmterm{DnaRegion} or \rbmterm{SmallMolecule}.\\
\rbmterm{bqbiol:is} & A term representing an individual type of an Agent entity, e.g. a protein entry from UniProt.\\
\rbmterm{bqbiol:isVersionOf} & A term representing the class type of an Agent entity, e.g. a SO term for a DNA-based agent.\\
\hline
\underline{Site declarations:}  \\
\rbmterm{rdf:type} & \rbmterm{Site} \\
\rbmterm{hasState} & Identifier for an internal state.\\
\rbmterm{bqbiol:isVersionOf} & A term representing the type of the site, e.g. A SO term for a nucleic acid-based site or an InterPro term for an amino acid-based site.\\
\hline
\underline{Internal state declarations:}\\
\rbmterm{rdf:type} & \rbmterm{State} \\
\rbmterm{bqbiol:is} & A term representing the state assignment, e.g. a term from the PSIMOD or the PO.\\
\hline
\underline{Rules:} \\
\rbmterm{rdf:type} & \rbmterm{Rule} \\
\rbmterm{dct:isPartOf} & Identifier for the \rbmterm{Model}.\\
\rbmterm{bqbiol:is} & A term representing an individual type of a rule, e.g. a KEGG entry.\\
\rbmterm{bqbiol:isVersionOf} & A term representing a class type of a rule, e.g. an EC number, a SO term or a GO term.\\
\rbmterm{subrule} & Identifier for a \rbmterm{Rule} entity.\\
\rbmterm{lhs}$^\dagger$ \rbmterm{rhs}$^\dagger$ & References to the patterns forming the left and right hand side of the rule.\\
\hline
\underline{Observables:}\\
\rbmterm{rdf:type} & \rbmterm{Observable} \\
\rbmterm{dct:isPartOf} & Identifier for the \rbmterm{Model}.\\
\rbmterm{pattern}$^\dagger$ & References the constituent patterns.\\
\hline
\underline{Patterns:} \\
\rbmterm{rdf:type} & \rbmterm{Pattern} \\
\rbmterm{ro:hasFunction} & A GO term specifying a biological function.\\
\rbmterm{agent}$^\dagger$ & Reference to the corresponding \rbmterm{Agent} declaration\\
\rbmterm{internalState}$^\dagger$ & Reference to a representation of a site's state\\
\rbmterm{isStatusOf}$^\dagger$ & Reference from a site's state to the corresponding site\\
\hline
\underline{Variables:}\\
\rbmterm{rdf:type} & \rbmterm{sbo:SBO:0000002} (\textit{quantitative systems description parameter})\\ 
\rbmterm{dct:isPartOf} & Identifier for the \rbmterm{Model}.\\
\rbmterm{bqbiol:isVersionOf} & A term representing a variable type. If exists, the term should a subterm of \rbmterm{SBO:0000002}.\\
\end{tabular}}
\end{table}

\end{document}